\documentclass[showpacs,preprintnumbers,superscriptaddress,nofootinbib,aps,onecolumn]{revtex4}
\usepackage{amsmath}
\usepackage{portland}
\usepackage[dvips]{graphicx}
\usepackage{dcolumn}
\usepackage{bm}
\usepackage{amsfonts}
\usepackage{amssymb}
\usepackage{scalefnt}
\usepackage[center]{caption2}

\input{tcilatex}

\begin{document}

\preprint{APS/123-}
\title{Backbending phenomena in light nuclei at $A\sim60$ mass region}
\author{S. U. El-Kameesy}
\email{saimrkhamisy@yahoo.com}
\affiliation{Department of Physics, Faculty of Science, Ain-Shams University, Cairo,
Egypt.}
\author{H. H. Alharbi }
\email{alharbi@kacst.edu.sa}
\affiliation{National Center for Mathematics and Physics, KACST, P.O. Box 6086, Riyadh
11442, Saudi Arabia,}
\author{H. A. Alhendi}
\email{alhendi@ksu.edu.sa}
\affiliation{Department of Physics and Astronomy, College of Science, King Saud
University, P.O. Box 2455, Riyadh 11454, Saudi Arabia}
\date{\today }

\begin{abstract}
Recent studies of the backbending phenomenon in medium light weight nuclei
near $A\sim 60$ expanded greatly our interest about how the single particle
orbits are nonlinearly affected by the collective motion. As a consequence
we have applied a modified version of the exponential model with the
inclusion of paring correlation to describe the energy spectra of the ground
state bands and/or the backbending phenomenon in mass region at $A\sim 60$.
A firm conclusion is obtained concerning the successful validity of the
proposed modified model in describing the backbending phenomenon in this
region. Comparison with different theoretical descriptions is discussed.
\end{abstract}

\maketitle

\section{INTRODUCTION}

Investigations of the ground state bands of nuclei at $A\sim60$ mass region
have recently become a particularly interesting subject in nuclear structure
studies \cite{Cameron,Brandolini,Caurier,Velazques,Hara}. These nuclei
exhibit a range of interesting features, including oblate and prolate
deformations as well as rapid variations in shape as a function of both spin
and particle number. The sudden disappearance of E2 strength at certain
spins indicates a shape change and requires the inclusion of upper pf
configuration \cite{Cameron2}. The shell model predictions have allowed
calculations in the full fp model space. These calculations have shown that
the collective properties of rotor like energies, backbending and large
B(E2) values can be reproduced.

Hara \textit{et al }\cite{Hara}\textit{.} studied the backbending mechanism
of $^{48}$Cr within the projected shell model (PSM) \cite{Hara2}, which has
been successful in describing well deformed heavy nuclei and those of
transitional region \cite{Sheikh}, it was concluded that the backbending in $%
^{48}$Cr is due to a band crossing involving an excited band built on
simultaneously broken pairs of neutrons and protons in the intruder subshell 
$f_{7/2}$. This result differs from that of Tanaka \textit{et al} \cite%
{Tanaka} based on the Cranked Hartee--Fock--Bogoliubov (CHFB), which claims
that the backbending in $^{48}$Cr is not due to level crossing.

\qquad The application of the generator coordinate method (GCM) has showed
that the backbending in $^{48}$Cr can be interpreted as due to crossing
between the deformed and spherical bands \cite{Hara2}. Accordingly, while
the backbending phenomenon in medium heavy nuclei are well described and
commonly understood as a band crossing phenomenon involving strong pairing
correlation \cite{Sorenson}, the origin of backbending in medium light
nuclei has been debated. Additionally, the role of the paring force in the
backbending phenomenon is not clearly outlined.

\qquad The interest of the $1f_{7/2}$ nuclei has been extended to levels
above the $1f_{7/2}$ band termination, in particular in connection with a
possible building up of superdeformation \cite{Lenzi}, where the quality of
SM calculations is probably not as good because of the possible
contributions from the SD shell and the $1g_{9/2}$ orbital \cite{Brandolini}%
. Cranking model analysis of $^{80}$Br energy levels reveals a signature
inversion at a spin of $12\hbar $ and a probably neutron alignment at $\hbar
\omega \approx $ $0.7$ MeV. The results are discussed within the framework
of the systematics of similar bands in the lighter Br isotopes and the
cranked--shell model \cite{Ray}.

\qquad The present study has been initiated due to the above contradictions
of different models in describing the backbending phenomenon in medium light
nuclei. Our study is based on applying a modified version of the exponential
model with pairing attenuation \cite{Alhendi}. It is hoped by such work to
have a strong evidence that the pairing force contribution still plays an
effective role in the backbending mechanism in this mass region at $A\sim 60$%
. In the following section we briefly present the model and in the next
section the model is applied to some even-even medium light nuclei. Finally
the last section contains our conclusion.

\section{Model Description}

\qquad Sood and Jain \cite{Sood} have previously developed an exponential
model based on the exponential dependence of the nuclear moment of inertia
on pairing correlation \cite{Bohr}. They gave the following relation:

\begin{equation}
E\left( I\right) =\frac{\hbar^{2}}{2\varphi_{0}}I\left( I+1\right) Exp\left[
\Delta_{0}\left( 1-\frac{I}{I_{c}}\right) ^{\frac{1}{2}}\right]  \label{Exp1}
\end{equation}

Excellent results have been obtained by means of this approach in describing
the ground state bands in deformed nuclei up to the point where backbending
occurs. They selected the value $18\hbar $ for $I_{C}$ as an input cutoff
corresponds to the point where the rotational frequency $\omega $ in the $%
\varphi -\omega ^{2}$ plots reaches a minimum value and the pairing
correlations disappear completely. Zhou and Zheng \cite{Celeghini} have
demonstrated that $I_{C}=$ $85\hbar $ is a suitable choice in their
calculations concerning superdeformed bands near $A\sim 190$, since the
pairing correlation in that region is still strong event at very high
rotational frequency. For medium light nuclei, $I_{C}$ can take values
smaller than $18\hbar $ because the backbending phenomenon in this region ($%
A $ $\approx $ $60$) lies at spin $I\approx $ $10\hbar $ \cite{Velazques}.
These works led us to use a suitable $I_{C}$ values to represent both the
variation of the moment of inertia and the paring correlation and to give
the model the ability to describe well the $\varphi -\omega ^{2}$ plot
regions, in particular the forward and down-bending regions, which lie after
the backbending region. The modified version of the exponential model with
pairing attenuation has the following form \cite{Alhendi,Alharbi}:

\begin{equation}
E\left( I\right) =\frac{\hbar^{2}}{2\varphi_{0}}I\left( I+1\right) Exp\left[
\Delta_{0}\left( 1-\frac{I}{I_{c}}\right) ^{\frac{1}{\nu}}\right]
\label{Exp2}
\end{equation}

Where $\varphi _{0}$, $\Delta _{0}$ and $\nu $ are the free parameters of
the model, which are adjusted to give a least-square fit to the experimental
data. This approach is supported by Ma and Rasmussen suggestion that there
is an exponential dependence of the moment of inertia on the parameter $\nu $
for a wide range $\nu $ values \cite{Ma}.

\section{Application to some even-even medium light nuclei}

The anomalous behavior, i.e. backbending of several medium light even-even
nuclei (Zn, Ge, Se, Kr, Sr), has been studied using our improved modified
version of the exponential model with pairing attenuation. The parameters of
the model (Table 1) were determined by means of a least-square fitting
procedure involving the experimental known energy levels \cite{iaea}.

\qquad The plots of the calculated data of $2\varphi _{I}/\hbar ^{2}$ versus
($\hbar \omega $)$^{2}$ for these isotopes are given in figure 1, where the
experimental data are also presented. From the excitation energies $E\left(
I\right) $ of the yrast bands we deduce the moment of inertia and the
squared rotational frequency $\omega ^{2}$ by using the well-known relations

\begin{equation}
\frac{2\varphi}{\hbar^{2}}=\frac{4I-2}{E\left( I\right) -E\left( I-2\right) }%
\text{ ,}  \label{phi}
\end{equation}

and

\begin{equation}
\left( \hbar\omega\right) ^{2}=\left( I^{2}-I+1\right) \left[ \frac{E\left(
I\right) -E\left( I-2\right) }{2I-1}\right] ^{2}  \label{hw}
\end{equation}

In figure 1 the experimental data show a clear evidence of backbending
phenomenon in all the presented nuclei at $I=8-12\hbar $. It is clear from
the same figure that the predictions of the applied improved exponential
model reproduce very well the backbending phenomenon in those nuclei and its
application improves as $A$ increases. This result may give an indication
that the pairing force contribution to the backbending phenomenon increases
as $A$ increases in the mass region under investigation. Another noticeable
success of the model is shown in Figure 1 \textquotedblleft concerning $^{68}
$Ge, $^{72}$Se, $^{78}$Kr and $^{80}$Sr\textquotedblright\ where the forward
and down-bending regions are very well described by its calculations.

\begin{table}[!htbp] \centering\scalefont{0.9}%
\caption
{The fitting parameters of the present model.}%
\begin{tabular}{ccccc}
\hline\hline
\textbf{Nucleus} & $\mathbf{2\varphi }_{0}\mathbf{/\hbar }^{2}$ & $\mathbf{%
\Delta }_{0}$ & $\mathbf{\nu }$ & $\mathbf{I}_{c}$ \\ \hline
$^{50}$Cr & 39.8033 & 1.66542 & 0.678977 & 26 \\ 
$^{62}$Zn & 16.1483 & 1.30409 & 0.313542 & 18 \\ 
$^{64}$Ge & 15.5511 & 1.38726 & 0.234353 & 18 \\ 
$^{68}$Ge & 25.516 & 1.90454 & 0.401219 & 20 \\ 
$^{74}$Kr & 47.5087 & 1.03889 & 0.428975 & 40 \\ 
$^{78}$Kr & 42.5699 & 1.31551 & 0.124932 & 80 \\ 
$^{80}$Kr & 35.0581 & 1.53869 & 0.452602 & 20 \\ 
$^{72}$Se & 38.7645 & 1.45564 & 0.375074 & 30 \\ 
$^{74}$Se & 40.435 & 1.26539 & 0.446761 & 28 \\ 
$^{76}$Se & 40.7577 & 1.60714 & 0.180245 & 50 \\ 
$^{78}$Se & 33.3453 & 1.50423 & 0.466925 & 20 \\ 
$^{78}$Sr & 40.4515 & 0.68491 & 0.125873 & 80 \\ 
$^{80}$Sr & 43.9731 & 1.03228 & 0.146307 & 80 \\ 
$^{82}$Sr & 48.9001 & 1.25575 & 0.173081 & 80 \\ \hline\hline
\end{tabular}%
\end{table}%

\bigskip \FRAME{ftbpFU}{3.9064in}{4.8879in}{0pt}{\Qcb{Calculated and
observed moment of inertia $2\protect\varphi /\hbar ^{2}$ vs. $(\hbar 
\protect\omega )^{2}$ for yrast levels of some light nuclei. The dots
represent experimental values.}}{}{greecefigure.eps}{\special{language
"Scientific Word";type "GRAPHIC";maintain-aspect-ratio TRUE;display
"USEDEF";valid_file "F";width 3.9064in;height 4.8879in;depth
0pt;original-width 7.7574in;original-height 11.0627in;cropleft "0";croptop
"0.9241";cropright "1";cropbottom "0.0453";filename
'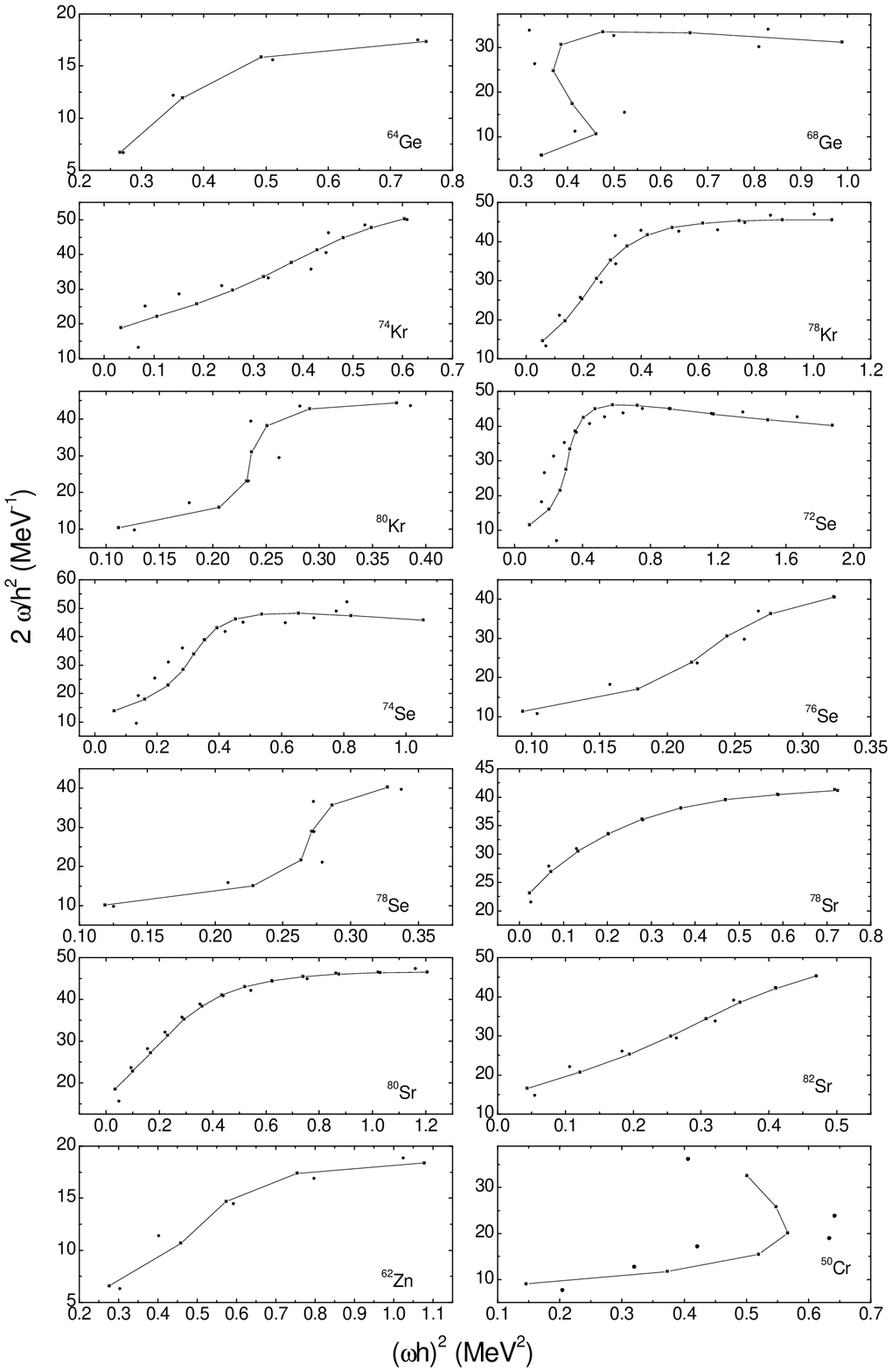';file-properties "XNPEU";}}

\section{\protect\bigskip Conclusion}

\qquad The present results of the improved exponential model with pairing
attenuation give a firm confirmation that the backbending in medium light
nuclei at low spins ($I=8-12\hbar $) can be interpreted due to paring force
which supports the band crossing mechanism in analogy with the earlier
calculations \cite{Hara} based on the projected shell model (PSM) and the
generator coordinator method (GCM). Furthermore, our simple modified formula
is able to describe well the forward and down-bending regions of the $%
\varphi -\omega ^{2}$ plots.

\end{document}